\documentclass[journal]{IEEEtran}

\makeatletter
\def\endthebibliography{%
	\def\@noitemerr{\@latex@warning{Empty `thebibliography' environment}}%
	\endlist
}
\makeatother

\newcommand{\RomanNumeralCaps}[1]
{\MakeUppercase{\romannumeral #1}}

\usepackage{cite}

\usepackage[pdftex]{graphicx} 

\usepackage[caption=false,font=footnotesize]{subfig}

\usepackage{amsmath}
\usepackage{amsfonts}
\usepackage{bm}

\usepackage{filecontents}
\usepackage{amssymb}
\usepackage{color}

\usepackage{epsfig}

\usepackage{verbatim}

\usepackage{url}

\usepackage{algorithm, tabularx}

\usepackage{lettrine}

\usepackage{lipsum}

\usepackage{siunitx}

\usepackage{soul,color}

\usepackage{mathrsfs}

\usepackage{array}

\usepackage[inline]{enumitem}

\usepackage{flushend}

\usepackage{dblfloatfix} 

\usepackage{algorithm}

\usepackage[noend]{algpseudocode}

\usepackage{setspace}
\usepackage{multirow}
\usepackage{array}
\newcolumntype{L}[1]{>{\raggedright\let\newline\\\arraybackslash\hspace{0pt}}m{#1}}
\newcolumntype{C}[1]{>{\centering\let\newline\\\arraybackslash\hspace{0pt}}m{#1}}
\newcolumntype{R}[1]{>{\raggedleft\let\newline\\\arraybackslash\hspace{0pt}}m{#1}}

\begin{document}
	
	\title{ Performance Analysis of Passive Retro-Reflector Based Tracking in Free-Space Optical Communications with Pointing Errors}

	
	\author{Hyung-Joo Moon,~\IEEEmembership{Graduate Student Member,~IEEE}, Chan-Byoung Chae,~\IEEEmembership{Fellow,~IEEE}, and \\Mohamed-Slim Alouini,~\IEEEmembership{Fellow,~IEEE}
		\thanks{This work was supported by the Institute of Information \& Communications Technology Planning \& Evaluation (IITP) funded by Korea government (MSIT) under Grants 2019-0-00685 and 2022-0-00704. (\textit{Corresponding author: C.-B. Chae.})}
		\thanks{H.-J. Moon and C.-B. Chae are with the School of Integrated Technology, Yonsei University, Seoul 03722, South Korea (e-mail: \{moonhj, cbchae\}@yonsei.ac.kr).}
		\thanks{M.-S. Alouini is with the Computer, Electrical, and Mathematical Science and Engineering Division, King Abdullah University of Science and Technology, Thuwal 23955, Saudi Arabia (e-mail: slim.alouini@kaust.edu.sa).}
	}
	
	{}
	

	\maketitle
	
	\begin{abstract}

In this correspondence, we propose a diversity-achieving retroreflector-based fine tracking system for free-space optical (FSO) communications. We show that multiple retroreflectors deployed around the communication telescope at the aerial vehicle save the payload capacity and enhance the outage performance of the fine tracking system. Through the analysis of the joint-pointing loss of the multiple retroreflectors, we derive the ordered moments of the received power. Our analysis can be further utilized for studies on multiple input multiple output (MIMO)-FSO. After the moment-based estimation of the received power distribution, we numerically analyze the outage performance. The greatest challenge of retroreflector-based FSO communication is a significant decrease in power.  Still,  our selected numerical results show that,  from an outage perspective, the proposed method can surpass conventional methods.

	\end{abstract}


	\IEEEpeerreviewmaketitle

	\section{Introduction}

\IEEEPARstart{F}{or} long-distance wireless communications with high capacity, free-space optical (FSO) communications has become one of the most promising communications technologies.  Unlike radio-frequency (RF) cellular communication networks, FSO communications are one-to-one due to the high directivity of laser beams. For precise beam pointing in FSO communications then, it is imperative to have a pointing, acquisition, and tracking (PAT) system~\cite{201701cst,2020wcnc}. The PAT system is divided into two steps--coarse pointing and fine tracking~\cite{eawcl}. At the initial stage, coarse pointing aims to achieve link availability, and, during the communication, fine tracking maintains the link from mechanical jitters and atmospheric turbulence.

A coarse pointing between the optical ground station (OGS) and the unmanned aerial vehicle (UAV) begins with the transmission of the UAV location information from the UAV to the OGS~\cite{eawcl}. Then the OGS transmits a beacon beam that covers the area where the UAV can exist. When the UAV receives the beacon beam, it aligns the pointing to the OGS and transmits the beam back to the incoming beam direction so that also the OGS can receive the beacon beam. When both sides are well-aligned through beacon beam reception, the fine tracking stage begins. During the fine tracking stage, the system requires more precise and fast compensation of pointing errors to keep both transceivers within the field of view. For this reason, quadrant detector (QD) and fast steering mirror (FSM) are widely used in this stage~\cite{2019ssc}. Based on the conventional fine tracking method using QD and FSM, we propose a fine tracking method that reduces outage probability and saves the power budget of the UAV.

In conventional fine tracking methods for two-way FSO communications, a beacon transmitter is deployed at both unmanned aerial vehicles (UAVs) and ground stations. In practice, however, the payload and power budget of UAVs are limited. We introduce a fine tracking method that replaces the beacon transmitter at the UAV with the multiple corner-cube reflectors (CCRs)--a device that reflects incident light in the same direction--to assist tracking at the ground station~\cite{200608spie}. There have been many studies on FSO communications in which a modulated retroreflector (MRR) replaces one side of the conventional FSO transceivers. In \cite{202106cl}, the authors analyze outage probability, average bit error rate (BER),  and ergodic capacity for the MRR-based FSO communications when nonzero boresight pointing error is assumed. The authors in~\cite{202103jlt}, test (through analysis and simulation) the feasiblity of the FSO communication using the micro CCR array.
Diffrent from previous studies, our proposed method assumes that the deployed CCRs are separated enough to achieve maximum path diversity. Also, we use passive CCRs to reflect a non-modulated beacon signal. Since each of the CCRs at the UAV sends the reflected beam back to the ground station, the received signal power is a sum of the uncorrelated reflected signals. This property allows the system to significantly reduce the link outage by achieving spatial diversity.
Additionally, a number of separated micro CCR arrays can replace CCRs for cost and weight reduction. However, we consider classical CCRs to avoid excessive assumptions and maintain mathematical simplicity.

In our proposed method, we base the methodology of the outage-performance analysis on the moment-matching approximation of the probability distribution function (PDF). The product of the uplink and downlink channel fading can be approximated as the $\alpha$-$\mu$ distribution~\cite{201808tcom} and the sum of the $\alpha$-$\mu$ distributed random variables (RVs) can also be approximated as the $\alpha$-$\mu$ distribution~\cite{200809twc, 202103twc}. Because of this, we approximate the sum power of reflected beams into the $\alpha$-$\mu$ distribution and derive the outage probability with a simple form of a cumulative distribution function (CDF). We further analyze the moment of the pointing-loss effect for the given deployment of a number of CCRs, which can be expanded into the pointing loss of the multiple input multiple output (MIMO)-FSO system.

The rest of this correspondence is organized as follows. In Section~\RomanNumeralCaps{2}, we introduce the signal model of the proposed retroreflector-based fine tracking system. We then describe the PDF of the pointing loss of an individual CCR. In Section~\RomanNumeralCaps{3}, we approximate the PDF of the received power at the ground station into the $\alpha$-$\mu$ distribution by the moment-matching method. Through this derivation, we present both exact and approximated moments. In Section~\RomanNumeralCaps{4}, we provide some selected simulation results, and we then finally provide our conclusions in Section~\RomanNumeralCaps{5}.

\section{System Model}

\subsection{Signal Power Model}
\label{powmod}
A conventional FSO channel model is as follows~\cite{200707jlt}:
\begin{equation}
\label{conv}
\begin{aligned}
P_{R}=h_{a}h_{\ell}h_{p}P_\text{T},
\end{aligned}
\end{equation}
where $P_{R}$ is a received power at the ground station, $h_{a}$, $h_{\ell}$, $h_{p}$, and $P_\text{T}$ denote channel fading, atmospheric loss, pointing loss, and transmit power at the UAV. Based on~(\ref{conv}), we formulate the signal power model for the proposed system model and describe the analytical characteristics of each term.

Assume that multiple CCRs are deployed around the communication telescope at the UAV; the reflected beacon signal power received at the ground station can be modeled as
\begin{equation}
\label{ccrsum}
\begin{aligned}
P_{\text{CCR}}=\sum_{i=1}^{M}P_\text{i},
\end{aligned}
\end{equation}
and the incoming signal power reflected from the $i$-th CCR is
\begin{equation}
\label{ccr}
\begin{aligned}
P_{i}=g_{a,i}g_{\ell}g_{p}\rho f_{a,i}f_{\ell}f_{p,i}P_\text{GS},
\end{aligned}
\end{equation}
where each of the parameters on the right-hand side indicates, respectively, downlink fading, downlink atmospheric loss, downlink pointing loss, reflection effect, uplink fading, uplink atmospheric loss, uplink pointing loss, and the transmit power of the ground station~\cite{201808tcom}.
We assume that the fading channels for different CCRs are independent~\cite{201108jocn} and fading channels of the uplink and downlink for each beam path are correlated. For further mathematical analysis, we substitute each term into the RV or a constant as follows:
\begin{equation}
\label{rv}
\begin{aligned}
X = f_a, Y = g_a, Z = f_p, c=g_{\ell}g_{p}\rho f_{\ell}P_\text{GS},
\end{aligned}
\end{equation}
\begin{equation}
\label{rvnum}
\begin{aligned}
X_i = f_{a,i}, Y_i = g_{a,i}, Z_i = f_{p,i}.
\end{aligned}
\end{equation}
The parameters $f_\ell$ and $g_\ell$ satisfies the Beer-Lambert law as~\cite{200402oe}
\begin{equation}
\label{attenuation}
\begin{aligned}
f_\ell, g_\ell = \exp(-\sigma{z}),
\end{aligned}
\end{equation}
where $z$ and $\sigma$ are a propagation distance and an attenuation coefficient, respectively. The size of the CCR determines the beam divergence of the reflected beam. Assume that the shape of the effective reflection area is a circle with a radius of $a_\text{Re}$\footnote[1]{The incident angle of the beam to the CCR affects the power of the reflected beam~\cite{197107ao}. However, we assume that the multiple CCRs are installed in the same direction on the quasi-static blimp. Thus, the effect of the variation in the incident angle is implied in $a_\text{Re}$, which is a constant.}, then the downlink beam divergence angle is determined as $\theta_\text{Re}=1.22\lambda/a_\text{Re}$ where $\lambda$ is a wavelength of the optical signal~\cite{197107ao}. Therefore, the value of $g_p$ is as follows:
\begin{equation}
\label{downpoint}
\begin{aligned}
g_p = 2a_\text{GS}^2/(z\theta_\text{Re})^2,
\end{aligned}
\end{equation}
where $a_\text{GS}$ is a radius of the ground station telescope. Since $\rho$ and $P_\text{GS}$ are the system parameters, $c$ in (\ref{rv}) is a constant and can be expressed as
\begin{equation}
\label{constant}
\begin{aligned}
c=\frac{1.34\,a_\text{GS}^2a_\text{Re}^2}{z^2\lambda^2}\exp({-2\sigma{z}}).
\end{aligned}
\end{equation}

\begin{figure}[t]
	\begin{center}
		{\includegraphics[width=0.8\columnwidth,keepaspectratio]
			{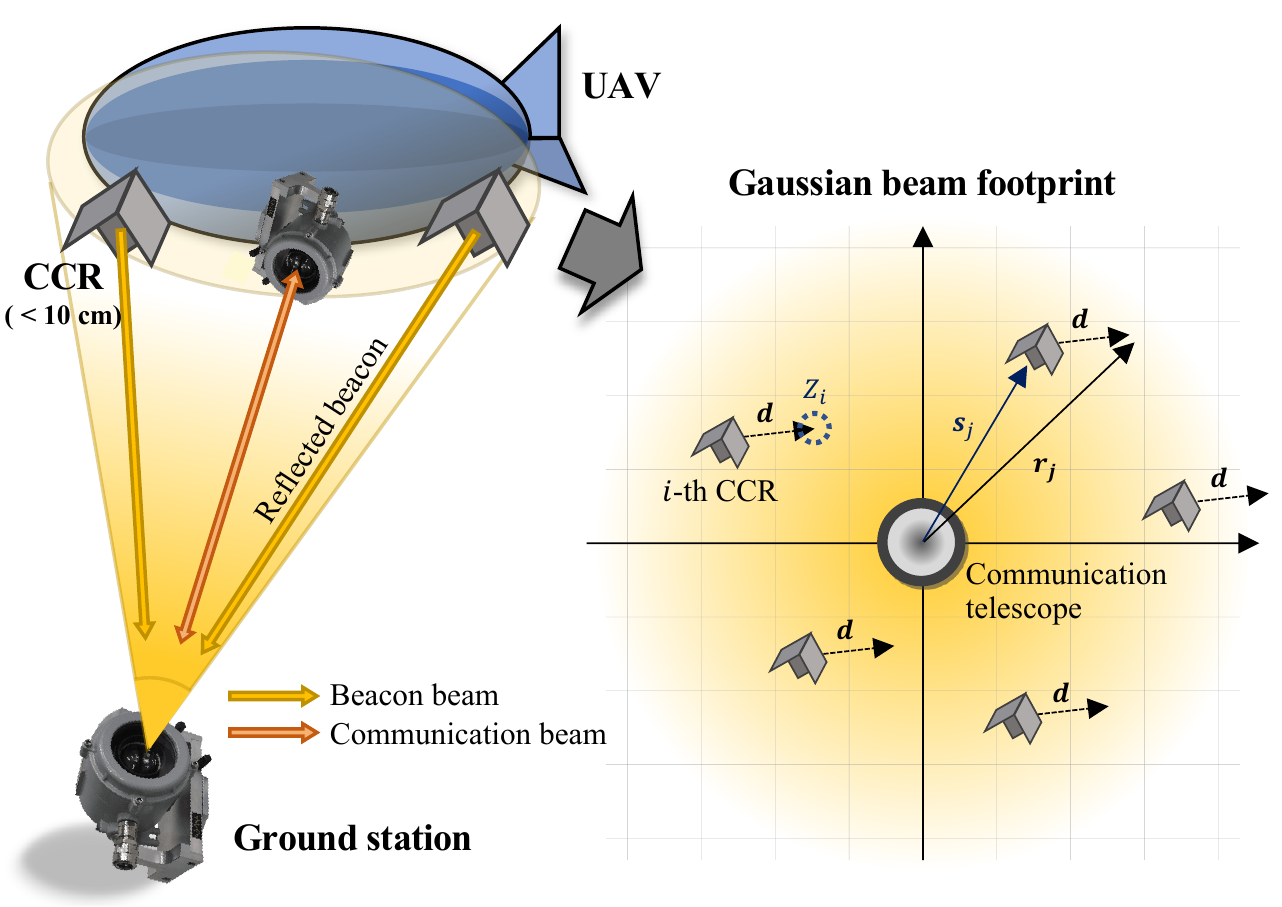}%
			\caption{In interpreting the joint-pointing loss, one must consider the deployment of CCRs on the beam footprint plane.}
			\label{main}
		}
	\end{center}
	\vspace{-15pt}
\end{figure}

Both $X_i$ and $Y_i$ follow the same Gamma-Gamma distribution for each $i$ and are correlated due to the channel reciprocity~\cite{200108oe}. Since Gamma-Gamma RV is a product of two uncorrelated Gamma RVs, the correlation coefficient is defined at this level. We can decompose the product of the uplink and downlink fading channel into four Gamma variables as
\begin{equation}
\label{decomposition}
\begin{aligned}
U = XY = {X^{(\alpha_1)}}{X^{(\beta_1)}}\cdot{Y^{(\alpha_2)}}{Y^{(\beta_2)}},
\end{aligned}
\end{equation}
where $\alpha_1$, $\beta_1$, $\alpha_2$, and $\beta_2$ are a unique parameter that determines the Gamma distribution. Because uplink and downlink have the same path at negligible time intervals, $\alpha_1 = \alpha_2$ and $\beta_1 = \beta_2$ can be assumed. Thus, the marginal PDF of ${X^{(\alpha_1)}}$ and ${Y^{(\alpha_2)}}$ are the same and can be expressed as follows:
\begin{equation}
\label{gammaalpha}
\begin{aligned}
f_\text{$\alpha_1$}(X^{(\alpha_1)})=\frac{\alpha_1(\alpha_1{x})^{\alpha_1-1}}{\Gamma(\alpha_1)}e^{-\alpha_1{x}},
\end{aligned}
\end{equation}
where $\Gamma(\cdot)$ is the Gamma function and the shape parameter and scale parameter are $\alpha_1$ and $1/\alpha_1$, respectively. Similarly, the marginal PDF of ${X^{(\beta_1)}}$ and ${Y^{(\beta_2)}}$ is
\begin{equation}
\label{gammabeta}
\begin{aligned}
f_\text{$\beta_1$}(X^{(\beta_1)})=\frac{\beta_1(\beta_1{x})^{\beta_1-1}}{\Gamma(\beta_1)}e^{-\beta_1{x}},
\end{aligned}
\end{equation}
where $\beta_1$ and $1/\beta_1$ are the parameters. Then the channel reciprocity is expressed by the channel correlation as $\rho_{\alpha}=\bold{corr}(X^{(\alpha_1)},Y^{(\alpha_2)})$ and $\rho_{\beta}=\bold{corr}(X^{(\beta_1)},Y^{(\beta_2)})$.
As each of the fading channels is indexed as $U_i = X_iY_i$, the entire randomness of $P_\text{CCR}$ can be described with the following RV:
\begin{equation}
\label{pccrrv}
\begin{aligned}
S=\sum_{i=1}^{M}U_iZ_i = \frac{P_\text{CCR}}{c}.
\end{aligned}
\end{equation}
The rest of the channel parameters are included in $c$ as~(\ref{rv}), which is a constant for every single CCR.

\subsection{PDF of $Z_i$}
As CCRs are distributed around the communication telescope (as shown in Fig.~\ref{main}), when analyzing the pointing loss $Z_i$, each CCR has a given boresight error. This can be described as the following system model. We define the position of the communication telescope as an origin of the two-dimensional coordinate plane. Then, the location of the CCR, beam displacement from the center point, and the superposition of two vectors can be defined, respectively, as follows:
\begin{equation}
\label{xy}
\begin{aligned}
\bold{s}_i=[s_{i,x}, s_{i,y}]^T, \bold{d}=[d_x, d_y]^T, \bold{r}_i=[r_{i,x}, r_{i,y}]^T.
\end{aligned}
\end{equation}
Assuming that both the incident beam and reflected beam are a Gaussian beam at the far field (see \cite[Sec. 4.5.2]{2005spie}), we arrive at
\begin{equation}
\label{xygaussianbeam}
\begin{aligned}
Z_i(\bold{r}_i;w)=A_0\exp{\left(-\frac{2|\bold{r}_i|^2}{w^2}\right)},
\end{aligned}
\end{equation}
where $w$ is a beamwidth, which follows $w=z\theta_\text{GS}$ for the uplink beam divergence angle $\theta_\text{GS}$ and $A_0={2a_\text{Re}^2}/{w^2}$~\cite{200707jlt}.
Since $\bold{d}$ is a beam displacement caused by the residual angle jitter of the fine tracking system, it follows a zero-mean multivariate normal distribution with the covariance matrix of $\Sigma_\bold{r}=\text{diag}(\sigma_s^2,\sigma_s^2)$. Thus, the PDF of $\bold{r}_i$ is
\begin{equation}
\label{xyrician}
\begin{aligned}
f_{\bold{r}_i}(\bold{r})=\frac{1}{2\pi{\sigma_s^2}}\exp{\left(-\frac{1}{2}(\bold{r}-\bold{s}_i)^T\Sigma_\bold{r}^{-1}(\bold{r}-\bold{s}_i)\right)},
\end{aligned}
\end{equation}
which then results in the following PDF~\cite{201402tcom}:
\begin{equation}
\label{fzdist}
\begin{aligned}
f_{Z_i}(Z)=\frac{w^2}{4\sigma_s^2}\cdot\frac{1}{Z}\bigg(\frac{Z}{A_0}\bigg)^{\frac{w^2}{4\sigma_s^2}}e^{-\frac{s_i^2}{2\sigma_s^2}}I_0\left(\frac{s_i}{\sigma_s^2}\sqrt{-\frac{w^2}{2}\ln\frac{Z}{A_0}}\right)\\
0\leq{Z}\leq{A_0},
\end{aligned}
\end{equation}
where $s_i=|\bold{s}_i|$ and $I_0(\cdot)$ is a modified Bessel function of the first kind of order zero.

\section{Outage Probability of Retroreflector Based Fine Tracking}
\label{sec4}

According to the system model, an outage probability of the received power can be defined as $\bold{Prob}[P_\text{CCR}<P_\text{th}]=\bold{Prob}[S<P_\text{th}/c]$. The RV $S$ is very complex, so that the derivation of an exact distribution is almost impossible. Hence, in this section, we derive the moments of $S$ and approximate the PDF into the $\alpha$-$\mu$ distribution by the moment-matching method.

\subsection{Moment Matching}

The PDF of the $\alpha$-$\mu$ RV $R$ is~\cite{200701tvt}
\begin{equation}
\label{alphamupdf}
\begin{aligned}
f_R(r)=\frac{\alpha\mu^{\mu}r^{\alpha\mu-1}}{\hat{r}^{\alpha\mu}\Gamma(\mu)}\exp\left(-\mu\frac{r^\alpha}{\hat{r}^\alpha}\right),
\end{aligned}
\end{equation}
where $\alpha>0$, $\mu = E[r^\alpha]^2/\text{Var}[r^\alpha]$, and $\hat{r} = {E[Z^\alpha]}^{\frac{1}{\alpha}}$. Its CDF is given by
\begin{equation}
\label{alphamucdf}
\begin{aligned}
F_R(r)=\frac{\Gamma(\mu,\mu{r^\alpha}/\hat{r}^\alpha)}{\Gamma(\mu)},
\end{aligned}
\end{equation}
where $\Gamma(z,y)=\int_0^yt^{z-1}\exp(-t)\,dt$ is the incomplete Gamma function.
To approximate $S$ into $R$, we use $1$st-, $2$nd-, and $4$th-order moments of two RVs for the moment-matching method. The $k$th-order moment of $R$ is~\cite{200809twc}
\begin{equation}
\label{alphamumoment}
\begin{aligned}
E[R^k]=\hat{r}^k\frac{\Gamma(\mu+k/\alpha)}{\mu^{k/\alpha}\Gamma(\mu)}.
\end{aligned}
\end{equation}
The reduced form of the moment-based estimators for $\alpha, \mu$, and $\hat{r}$ are as follows:
\begin{equation}
\label{momentest1}
\begin{aligned}
\frac{\Gamma^2(\mu+1/\alpha)}{\Gamma(\mu)\Gamma(\mu+2/\alpha)-\Gamma^2(\mu+1/\alpha)}=\frac{E^2[S]}{E[S^2]-E^2[S]},
\end{aligned}
\end{equation}
\begin{equation}
\label{momentest2}
\begin{aligned}
\frac{\Gamma^2(\mu+2/\alpha)}{\Gamma(\mu)\Gamma(\mu+4/\alpha)-\Gamma^2(\mu+2/\alpha)}=\frac{E^2[S^2]}{E[S^4]-E^2[S^2]},
\end{aligned}
\end{equation}
\begin{equation}
\label{momentest3}
\begin{aligned}
\hat{r}=\frac{\mu^{1/\alpha}\Gamma(\mu)E[S]}{\Gamma(\mu+1/\alpha)}.
\end{aligned}
\end{equation}

In order to solve~(\ref{momentest1}), (\ref{momentest2}), and (\ref{momentest3}), we then have to derive $1$st-, $2$nd-, and $4$th-order moments of $S$. The $n_0$th-order moment of $S$ can be developed as
\begin{equation}
\label{netmoment}
\begin{aligned}
E[S^{n_0}]=\sum_{n_1=0}^{n_0}\sum_{n_2=0}^{n_1}&\cdots\sum_{n_{M-1}=0}^{n_{M-2}}{\binom{n_0}{n_1}\binom{n_1}{n_2}\cdots\binom{n_{M-2}}{n_{M-1}}}\\
&{\cdot}E[U_1^{n_0-n_1}]E[U_2^{n_1-n_2}]{\cdots}E[U_M^{n_{M-1}}]\\
&{\cdot}E[Z_1^{n_0-n_1}Z_2^{n_1-n_2}{\cdots}Z_M^{n_{M-1}}]
\end{aligned}
\end{equation}
from (\ref{pccrrv}). By~(\ref{decomposition}), we can express the ordered moments of $U$ as follows~\cite{200506el}:
\begin{equation}
\label{moment}
\begin{aligned}
E[{U^n}]=&\frac{\Gamma(\alpha_1+n)^2\Gamma(\beta_1+n)^2}{\Gamma(\alpha_1)^2\Gamma(\beta_1)^2}(\alpha_1\beta_1)^{-2n}\\&\cdot{}_2F_1(-n,-n;\alpha_1;\rho_\alpha){}_2F_1(-n,-n;\beta_1;\rho_\beta),
\end{aligned}
\end{equation}
where $_pF_q(\cdot)$ is the generalized hypergeometric function. To calculate the joint-ordered moments of $Z_i$s, we derive the exact and approximated form of $E[Z_1^{n_0-n_1}Z_2^{n_1-n_2}{\cdots}Z_M^{n_{M-1}}]$. For convenience, we transform the formula as follows:
\begin{equation}
\label{momenttransform}
\begin{aligned}
E[Z_1^{n_0-n_1}Z_2^{n_1-n_2}{\cdots}Z_M^{n_{M-1}}]=E[Z_{m_1}Z_{m_2}{\cdots}Z_{m_{n_0}}],
\end{aligned}
\end{equation}
where $m_1=\cdots=m_{n_{M-1}}=M$, $m_{n_{M-1}+1}=\cdots=m_{n_{M-2}}=M-1$, $\cdots$, $m_{n_{1}+1}=\cdots=m_{n_{0}}=1$.

Starting from the following equation:
\begin{equation}
\label{jointmomentequation}
\begin{aligned}
&E[Z_{m_1}Z_{m_2}{\cdots}Z_{m_{n_0}}]=\int_0^{2\pi}\int_0^{\infty}\prod_{i=1}^{n_0}Z_{m_i}\cdot\frac{\delta}{2\pi{\sigma_s^2}}e^{-\frac{\delta^2}{2\sigma_s^2}}\,d{\delta}\,d{\theta},
\end{aligned}
\end{equation}
where $\delta=|\bold{d}|$ and $\theta=\arg(\bold{d})$, we derive the exact moment including an integral operation and the approximated moment including combinatory sums of polynomials.

\subsection{Exact Moment}
\label{exactm}

\newtheorem{theorem}{Theorem}
\begin{theorem}{The exact form of~(\ref{jointmomentequation}) can be derived as
\begin{equation}
\label{jointmomentexact}
\begin{aligned}
&E[Z_{m_1}Z_{m_2}{\cdots}Z_{m_{n_0}}]\\
&=A_0^{n_0}e^{-\sum_{i=1}^{n_0}\frac{{2s}_{m_{i}}^2}{w^2}}\int_{0}^{\infty}e^{-\big(\frac{2n_0}{w^2}+\frac{1}{2\sigma_s^2}\big)\delta^2}\frac{\delta}{\sigma_s^2}I_{0}(K\delta)\,d\delta,
\end{aligned}
\end{equation}
where $K=\sqrt{\left(\sum_{i=1}^{n_0}\frac{4s_{m_i}\sin\phi_{m_i}}{w^2}\right)^2+\left(\sum_{i=1}^{n_0}\frac{4s_{m_i}\cos\phi_{m_i}}{w^2}\right)^2}$ and $\phi_i=\arg({\bold{s}_i})$.
}
\label{theo1}
\end{theorem}

\def\QEDmark{\ensuremath{\blacksquare}}
\def\proof{\emph{Proof: }}
\def\endproof{\hfill\QEDmark}

\proof
See Appendix~\ref{appen1}.
\endproof

\begin{figure}[t]
	\begin{center}
		{\includegraphics[width=0.9\columnwidth,keepaspectratio]
			{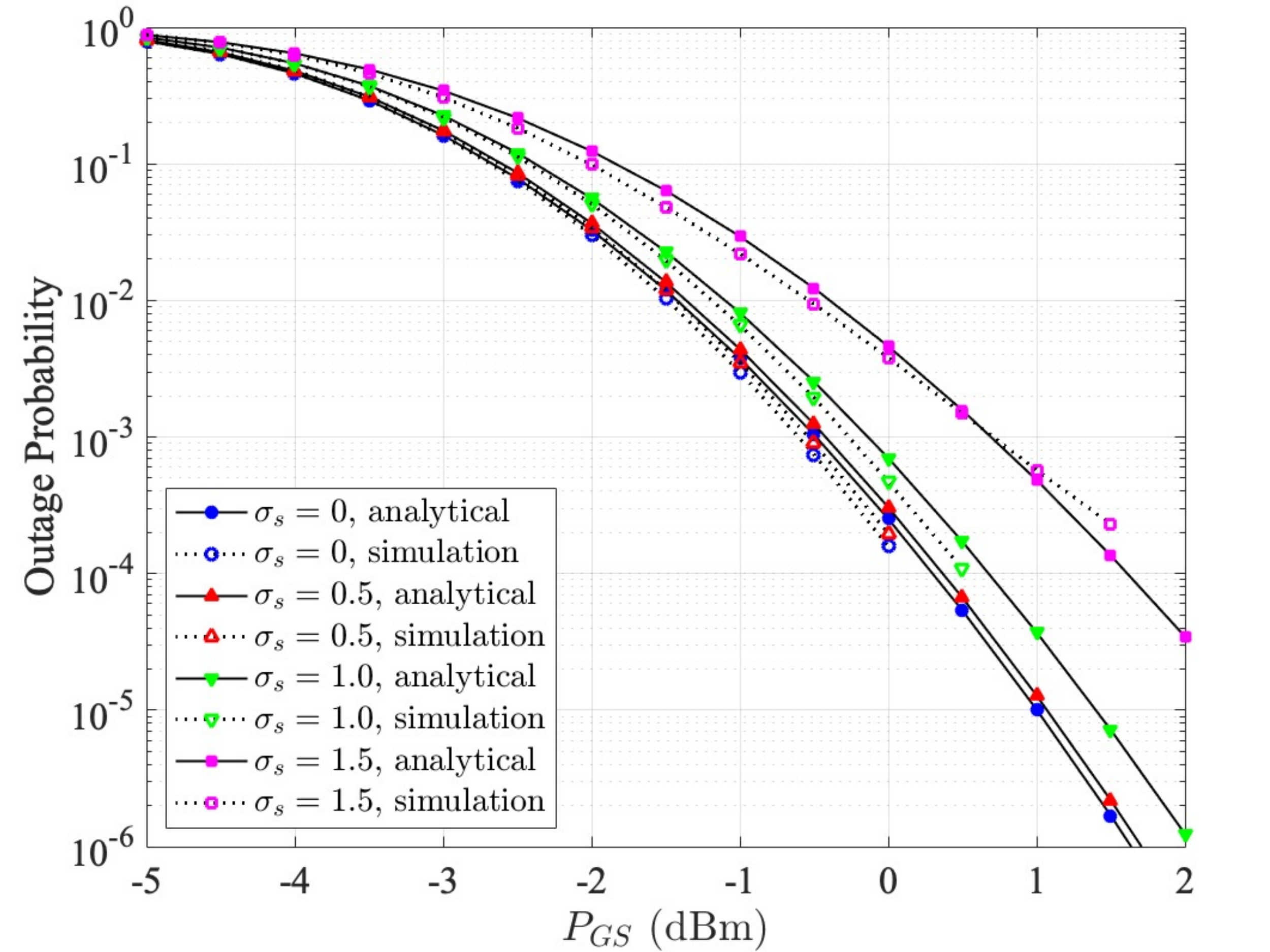}%
			\caption{Outage probability of the proposed fine tracking system in weak turbulence channels with $M=4$ and $w\approx8.5$ for different values of $\sigma_s$.}
			\label{ccr1}
		}
	\end{center}
	\vspace{-12pt}
\end{figure}

\begin{table}[!t]
	\centering
	\caption{Simulation parameters}
	\small
	\label{tbl}
	\begin{tabular}{p{5.5cm} c}
	\hline
	\bfseries{Parameter} 				& \bfseries{Value} \\ 
	\hline
	Visibility range ($V$)	& $10\,\,\si{\kilo\metre}$\\
	Link distance ($z$)	& $5\,\,\si{\kilo\metre}$\\
	Optical threshold power ($P_\text{th}$)	& $10\,\,\si{\nano\watt}$\\
	Radius of CCR ($a_\text{Re}$)	& $5\,\,\si{\centi\meter}$\\
	Radius of OGS telescope ($a_\text{GS}$)	& $10\,\,\si{\centi\meter}$\\
	Reflection effect ($\rho$)	& $0.5$\\
	Weak turbulence ($\alpha, \beta$)	& $17.1,\,\,16.0$\\
	Strong turbulence ($\alpha, \beta$)	& $4.0,\,\,1.9$\\
	Correlation coefficient ($\rho_\alpha, \rho_\beta$)	& $0.7$\\
	\hline
	\end{tabular} 
	\vspace{-10pt}
\end{table}

\begin{figure}[t]
	\begin{center}
		{\includegraphics[width=0.9\columnwidth,keepaspectratio]
			{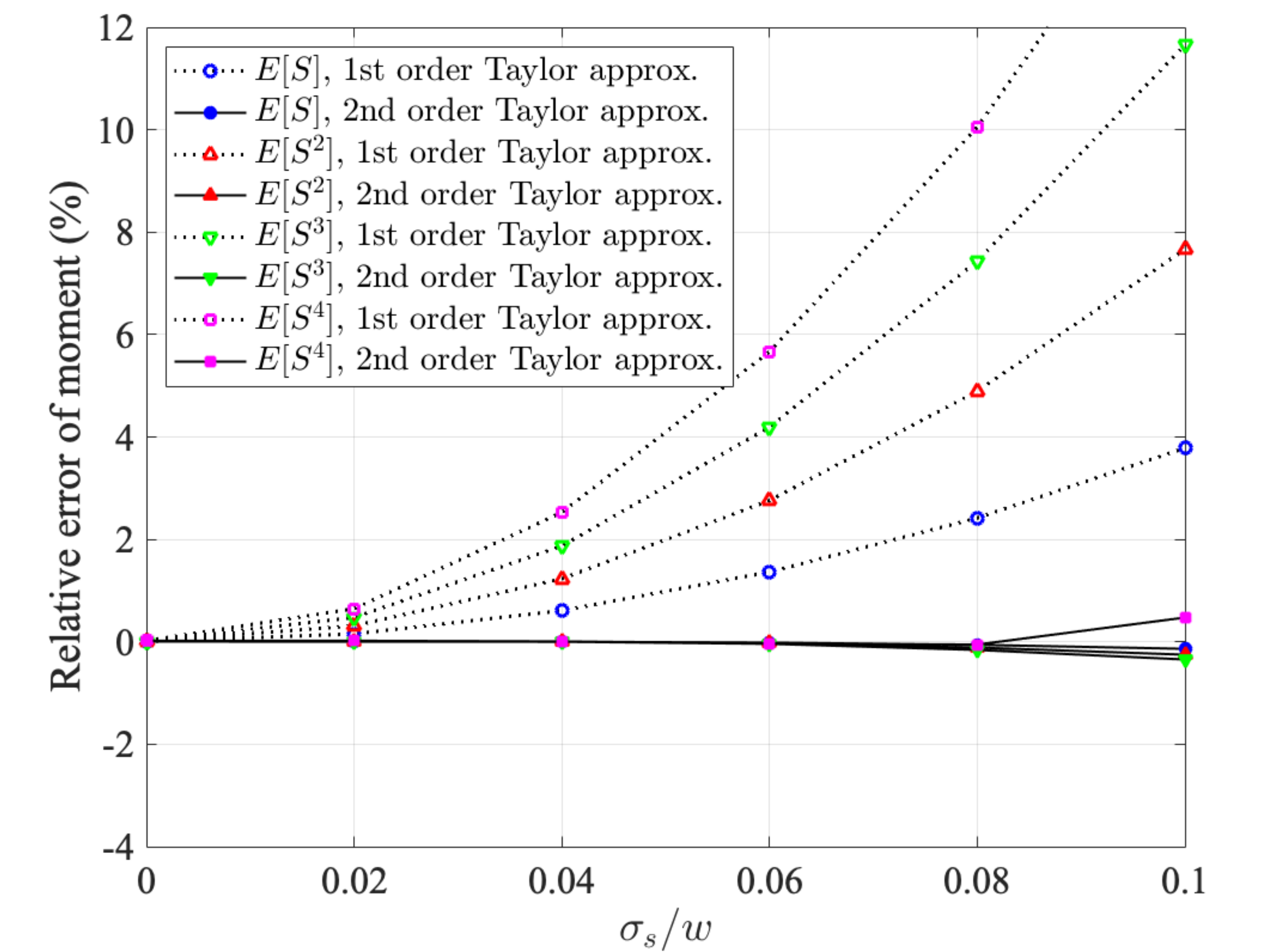}%
			\caption{Relative error of the moments of $S$ with respect to the ratio of $\sigma_s$ to $w$ with $M=4$ for different order of moments.}
			\label{ccr2}
		}
	\end{center}
	\vspace{-12pt}
\end{figure}

\subsection{Approximated Moment}
\label{approxm}

\begin{theorem}{The approximated form of~(\ref{jointmomentequation}) can be derived as
\begin{equation}
\label{jointmoment}
\begin{aligned}
&E[Z_{m_1}Z_{m_2}{\cdots}Z_{m_{n_0}}]\\
&=\frac{A_0^{n_0}}{2\pi}e^{-\sum_{i=1}^{n_0}\frac{2s_{m_i}^2}{w^2}}\sum_{\nu=1}^{n_0}\mu_{\sigma_s}^{(2\nu)}P_{n_0}^{(2\nu)}(m_1,m_2,\cdots,m_{n_0}),
\end{aligned}
\end{equation}\noindent
\begin{figure*}[b!]
\vspace{-10pt}
\begin{center}
\line(1,0){515}
\end{center}
\vspace{-6pt}
\begin{subequations}
\small
\label{jointmomentpolys}
\begin{align}
\begin{split}
\label{jointmomentpoly0}
P_{n_0}^{(0)}=&2\pi
\end{split}\\[-2pt]
\begin{split}
\label{jointmomentpoly2}
P_{n_0}^{(2)}=&\sum_{\substack{\text{Sym}\{k_i\}_{i=1}^2\in\mathcal{M}\\{\forall}k_p{\neq}k_q}}\frac{1}{2!}\prod_{j=1}^{2}\left(-\frac{4}{w^2}s_{m_{k_j}}\right)\mathcal{C}(\phi_{m_{k_1}},\phi_{m_{k_2}})+\sum_{\text{Sym}\,k_1\in\mathcal{M}}\bigg[\frac{8}{w^4}s_{m_{k_1}}^{2}\mathcal{C}(\phi_{m_{k_1}},\phi_{m_{k_1}})-\frac{4\pi}{w^2}\bigg]
\end{split}\\[-7pt]
\begin{split}
\label{jointmomentpoly4}
P_{n_0}^{(4)}=&\sum_{\substack{\text{Sym}\{k_i\}_{i=1}^4\in\mathcal{M}\\{\forall}k_p{\neq}k_q}}\frac{1}{4!}\prod_{j=1}^{4}\left(-\frac{4}{w^2}s_{m_{k_j}}\right)\mathcal{C}(\phi_{m_{k_1}},\phi_{m_{k_2}},\phi_{m_{k_3}},\phi_{m_{k_4}})\\[-7pt]
&+\sum_{\substack{\text{Sym}\{k_i\}_{i=1}^3\in\mathcal{M}\\{\forall}k_p{\neq}k_q}}\frac{1}{2!}\prod_{j=1}^{2}\left(-\frac{4}{w^2}s_{m_{k_j}}\right)\left(\frac{8}{w^4}s_{m_{k_3}}^{2}\mathcal{C}(\phi_{m_{k_1}},\phi_{m_{k_2}},\phi_{m_{k_3}},\phi_{m_{k_3}})-\frac{2}{w^2}\mathcal{C}(\phi_{m_{k_1}},\phi_{m_{k_2}})\right)\\[-7pt]
&+\sum_{\substack{\text{Sym}\{k_i\}_{i=1}^2\in\mathcal{M}\\{\forall}k_p{\neq}k_q}}\Bigg[\frac{1}{2!}\prod_{j=1}^{2}\left(\frac{8}{w^4}s_{m_{k_j}}^2\right)\mathcal{C}(\phi_{m_{k_1}},\phi_{m_{k_1}},\phi_{m_{k_2}},\phi_{m_{k_2}})-\frac{16}{w^6}s_{m_{k_1}}^2\mathcal{C}(\phi_{m_{k_1}},\phi_{m_{k_1}})+\frac{1}{2!}\frac{8\pi}{w^4}\Bigg]
\end{split}\\[-8pt]
\begin{split}
\label{jointmomentpoly6}
P_{n_0}^{(6)}=&\sum_{\substack{\text{Sym}\{k_i\}_{i=1}^4\in\mathcal{M}\\{\forall}k_p{\neq}k_q}}\frac{1}{2!}\prod_{j=1}^{2}\left(-\frac{4}{w^2}s_{m_{k_j}}\right)\bigg\{\frac{1}{2!}\prod_{j=3}^{4}\left(\frac{8}{w^4}s_{m_{k_j}}^2\right)\mathcal{C}(\phi_{m_{k_1}},\phi_{m_{k_2}},\phi_{m_{k_3}},\phi_{m_{k_3}},\phi_{m_{k_4}},\phi_{m_{k_4}})\\[-7pt]
&{\qquad\qquad\qquad\qquad\qquad\qquad\qquad\qquad}-\frac{16}{w^6}s_{m_{k_3}}^2\mathcal{C}(\phi_{m_{k_1}},\phi_{m_{k_2}},\phi_{m_{k_3}},\phi_{m_{k_3}})+\frac{1}{2!}\frac{4}{w^4}\mathcal{C}(\phi_{m_{k_1}},\phi_{m_{k_2}})\bigg\}\\[-1pt]
&+\sum_{\substack{\text{Sym}\{k_i\}_{i=1}^3\in\mathcal{M}\\{\forall}k_p{\neq}k_q}}\Bigg[\frac{1}{3!}\prod_{j=1}^{3}\left(\frac{8}{w^4}s_{m_{k_j}}^2\right)\mathcal{C}(\phi_{m_{k_1}},\phi_{m_{k_1}},\phi_{m_{k_2}},\phi_{m_{k_2}},\phi_{m_{k_3}},\phi_{m_{k_3}})+\frac{1}{2!}\prod_{j=1}^{2}\left(\frac{8}{w^4}s_{m_{k_j}}^2\right)\left(-\frac{2}{w^2}\right)\\[-11pt]
&{\qquad\qquad\qquad\qquad\qquad\qquad\qquad}\cdot\mathcal{C}(\phi_{m_{k_1}},\phi_{m_{k_1}},\phi_{m_{k_2}},\phi_{m_{k_2}})+\frac{1}{2!}\frac{32}{w^8}s_{m_{k_1}}^2\mathcal{C}(\phi_{m_{k_1}},\phi_{m_{k_1}})+\frac{1}{3!}\frac{16\pi}{w^6}\Bigg]
\end{split}\\[-5pt]
\begin{split}
\label{jointmomentpoly8}
P_{n_0}^{(8)}=&\sum_{\substack{\text{Sym}\{k_i\}_{i=1}^4\in\mathcal{M}\\{\forall}k_p{\neq}k_q}}\Bigg[\frac{1}{4!}\prod_{j=1}^{4}\left(\frac{8}{w^2}s_{m_{k_j}}^2\right)\mathcal{C}(\phi_{m_{k_1}},\phi_{m_{k_1}},\phi_{m_{k_2}},\phi_{m_{k_2}},\phi_{m_{k_3}},\phi_{m_{k_3}},\phi_{m_{k_4}},\phi_{m_{k_4}})\\[-9pt]
&{\qquad\qquad\quad\,}+\frac{1}{3!}\prod_{j=1}^{3}\left(\frac{8}{w^2}s_{m_{k_j}}^2\right)\left(-\frac{2}{w^2}\right)\mathcal{C}(\phi_{m_{k_1}},\phi_{m_{k_1}},\phi_{m_{k_2}},\phi_{m_{k_2}},\phi_{m_{k_3}},\phi_{m_{k_3}})\\[-6pt]
&{\qquad}+\frac{1}{2!2!}\prod_{j=1}^{2}\left(\frac{8}{w^2}s_{m_{k_j}}^2\right)\frac{4}{w^4}\mathcal{C}(\phi_{m_{k_1}},\phi_{m_{k_1}},\phi_{m_{k_2}},\phi_{m_{k_2}})-\frac{1}{3!}\frac{64}{w^{10}}s_{m_{k_1}}^2\mathcal{C}(\phi_{m_{k_1}},\phi_{m_{k_1}})+2\pi\frac{1}{4!}\frac{16}{w^8}\Bigg]
\end{split}
\end{align}
\end{subequations}\noindent
\end{figure*}\noindent
where $P_{n_0}^{(2\nu)}(m_1,m_2,\cdots,m_{n_0})$ can be developed as~(\ref{jointmomentpolys}) for $n_0\leq4$, $\phi_i=\arg({\bold{s}_i})$, and $\mathcal{M}=\{1,2,\cdots,n_0\}$.
A symbol $\mu_{\sigma_s}^{(2\nu)}$ is a $2\nu$th-moment of the Rayleigh distribution and has a value of
\begin{equation}
\label{rayleighmoment}
\begin{aligned}
\mu_{\sigma_s}^{(2\nu)}=2^{\nu}\nu!\,\sigma_s^{2\nu}.
\end{aligned}
\end{equation}
A function $\mathcal{C}(\cdot)$ is a definite integral of a product of cosine functions and can be organized into the sum of cosine functions as
\begin{equation}
\label{cosintegral}
\begin{aligned}
\mathcal{C}(\eta_1,\cdots,\eta_{2\ell})&=\int_{0}^{2\pi}\prod_{i=1}^{2\ell}\cos(\theta-\eta_{i})\,d\theta\\
&=\frac{2\pi}{2^{2\ell}(\ell!)^2}\sum_{\substack{\text{Sym}\{k_i\}_{i=1}^{2\ell}\in\mathcal{Z}\\
{\forall}k_p{\neq}k_q}}\cos\Bigg(\sum_{j=1}^{\ell}\eta_j-\eta_{\ell+j}\Bigg),
\end{aligned}
\end{equation}
where $\mathcal{Z}=\{1,2,\cdots,2\ell\}$.
}
\label{theo2}
\end{theorem}

\def\QEDmark{\ensuremath{\blacksquare}}
\def\proof{\emph{Proof: }}
\def\endproof{\hfill\QEDmark}

\proof
See Appendix~\ref{appen2}.
\endproof

\begin{figure}[t]
	\begin{center}
		{\includegraphics[width=0.9\columnwidth,keepaspectratio]
			{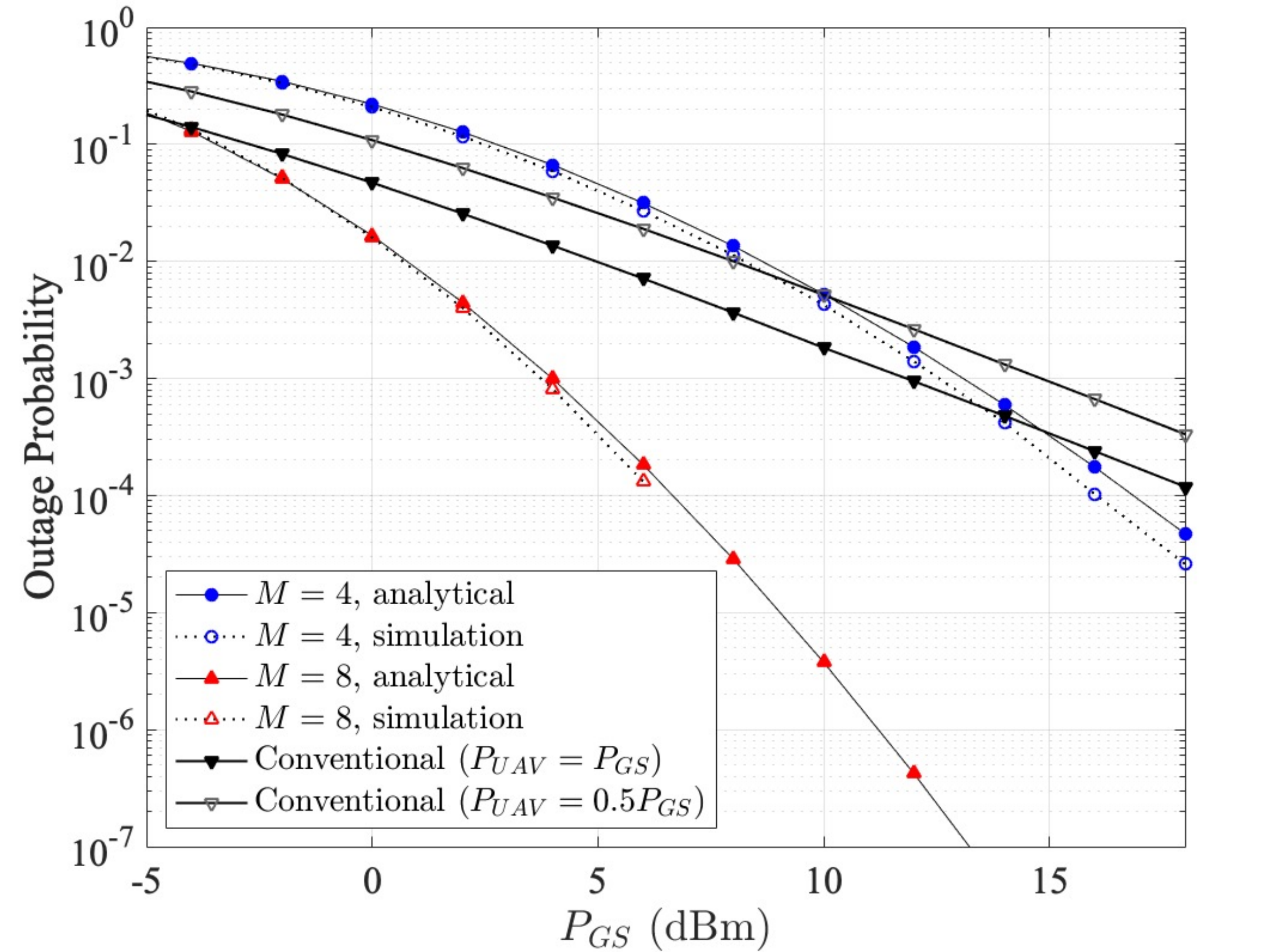}%
			\caption{Comparison between the outage probability of the proposed and conventional fine tracking systems in strong turbulence channel with $w=10$.}
			\label{ccr3}
		}
	\end{center}
	\vspace{-12pt}
\end{figure}

\begin{figure}[t]
	\begin{center}
		{\includegraphics[width=0.9\columnwidth,keepaspectratio]
			{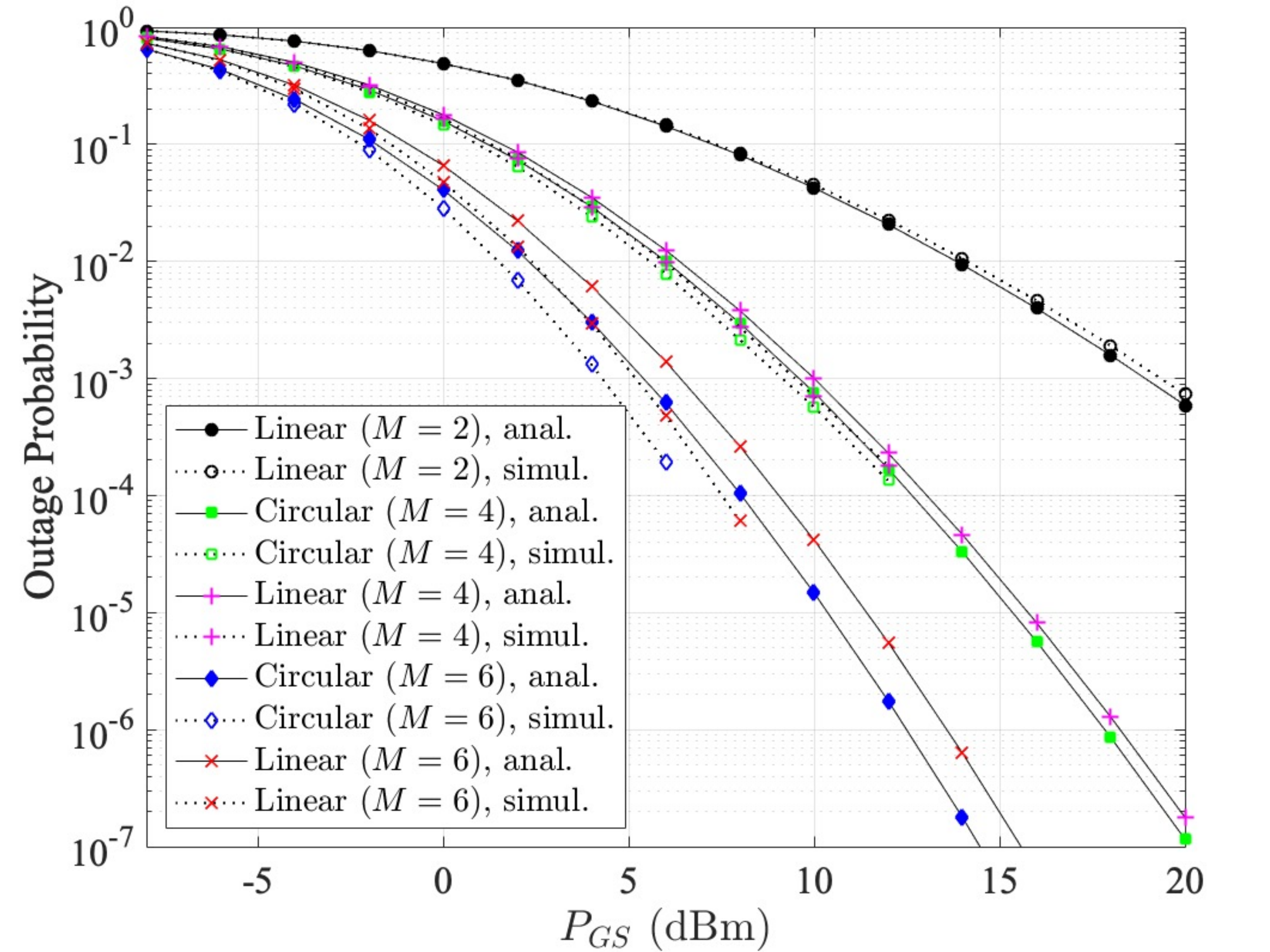}%
			\caption{Outage probability of the proposed fine tracking system in moderate channel with $w=10$ for different numbers and deployments of CCRs.}
			\label{ccr4}
		}
	\end{center}
	\vspace{-12pt}
\end{figure}

\section{Numerical Results}
\label{sec5}

In this section, we first discuss the implementation issues and the simulation parameter settings. Then, we show numerical results of the outage probability during the fine tracking stage. Table.~\ref{tbl} lists general simulation parameter values throughout this section. The link distance in the simulation is $5$~km, which can be considered as the altitude of the UAV\footnote[2]{The vertical link distance of $5$~km is grounded to the airspace Class~E in the United States, an altitude of $370$~m to $5500$~m. Through the simulations, we show that the proposed method is applicable to UAVs at the highest altitude of the airspace Class~E and below.}. For the proposed method, the link distance affects the received signal power by the atmospheric loss and free-space path loss (in (\ref{constant}) and (\ref{xygaussianbeam}), respectively) twice for the uplink and downlink. However, for the conventional method, the link distance only affects the downlink channels. Thus, the decreased link distance is always more advantageous to the proposed method than the conventional one. For this reason, the proposed method will perform better than the following outage results for the UAVs lower than the altitude of $5$~km.

The radius of the CCRs is set to $5$~cm, which is generally a larger size than most commercial passive CCRs. Considering the weight and size of the CCRs, we assume the blimp UAV to ensure sufficient CCR spacing and large payload capacity. That being said, the system providers can take advantage of the decreased operational altitude by launching smaller CCRs, which will considerably reduce the payload weight and operating costs. In this case, smaller UAVs, such as rotary-wing drones, can also carry multiple CCRs to apply our method. As noted in Sec.~\ref{powmod}, we assume that all the CCRs and the communication telescope are at least $\sqrt{2}$~m apart to preserve the channel independence\footnote[3]{According to~\cite{201108jocn}, atmospheric correlation length is about $59$~cm for the link distance of $5$~km and weak turbulence conditions. The weak turbulence is expressed by the refractive index structure constant,  as $C_n^2=10^{-17}$. In the simulation, the minimum CCR spacing is $\sqrt{2}$~m, which is larger than the correlation length.}~\cite{201108jocn}. CCRs in a linear deployment are aligned at equal intervals along the axis, and those in a circular deployment are listed at equal intervals above the circumference of radius $\sqrt{2}$~m. The moment-based parameter estimation of~(\ref{momentest1}), (\ref{momentest2}), and~(\ref{momentest3}) is calculated by the \textit{fsolve} function in MATLAB.
Moreover, the outage probability is obtained by~(\ref{alphamucdf}), with the estimated parameters.

As shown in Fig.~\ref{ccr1}, for different $\sigma_s$ values, the analytical results follow the simulation results, due to the joint-pointing loss derived in this paper.
In Fig.~\ref{ccr2}, we show the approximation error of~(\ref{jointmoment}), the moment of joint-pointing loss. As a point of comparison with our results, we also offer moments to which the $1$st-order Taylor approximation is applied.
In Fig.~\ref{ccr3}, we emphasize the diversity effect of multiple passive CCRs by comparing the outage probability of the proposed system to that of the conventional fine tracking system, where a beacon transmitter is used at aerial vehicles. In this case, we assume that the transmit power is equal to or half of the power at the ground station due to the limitation of the aerial payload. Furthermore, since we derived the joint-pointing loss for the given locations of CCRs, we compare (in Fig.~\ref{ccr4}) the outage performance of the systems with different CCR deployments around the communication telescope.

\section{Conclusion}
\label{sec6}

In this correspondence, we introduced and analyzed a novel, fine tracking system that uses multiple passive corner-cube reflectors (CCRs) for spatial diversity and power saving. For the system model in which a number of passive CCRs are distributed around the communication telescope at the aircraft, we formulated a received power model at the ground station. We then derived the exact and approximated moments to approximate the PDF into the $\alpha$-$\mu$ distribution. While a concern has been the low power of the reflected beam, the simulation results and analytical results support our argument that multiple passive CCRs can exceed the outage performance of the conventional method.

\vspace{-8pt}

\appendices
\section{Proof of Theorem~\ref{theo1}}
\label{appen1}
From~(\ref{xygaussianbeam}) and (\ref{jointmomentequation}), we obtain
\begin{equation}
\label{jointmomentexactproof}
\begin{aligned}
E&[Z_{m_1}Z_{m_2}{\cdots}Z_{m_{n_0}}]\\
=&A_0^{n_0}e^{-\sum_{i=1}^{n_0}\frac{{2s}_{m_{i}}^2}{w^2}}\\
&\int_{0}^{\infty}e^{-\big(\frac{2n_0}{w^2}+\frac{1}{2\sigma_s^2}\big)\delta^2}\frac{\delta}{2\pi\sigma_s^2}\int_{0}^{2\pi}e^{-\sum_{i=1}^{n_0}\frac{4s_{m_i}\delta\cos{(\phi_{m_i}-\theta)}}{w^2}}\,d\theta\,d\delta.
\end{aligned}
\end{equation}
Since the sum of cosine functions ${-\sum_{i=1}^{n_0}\frac{4s_{m_i}\delta\cos{(\phi_{m_i}-\theta)}}{w^2}}$ can be simplified into a single cosine function, the inner integral is then expressed as a modified Bessel function of the first kind. Then~(\ref{jointmomentexactproof}) results in~(\ref{jointmomentexact}).

\vspace{-8pt}

\section{Proof of Theorem~\ref{theo2}}
\label{appen2}

By substituting $\bold{r}=\bold{s}+\bold{d}$ into~(\ref{xygaussianbeam}) and (\ref{xyrician}) and applying $2$nd order Taylor approximation, the Gaussian beam profile at $\bold{s}$ results in an approximated form of $Z_i$ as
\begin{equation}
\label{dbeamapprox}
\begin{aligned}
Z_i{\approx}A_0&e^{-\frac{2|\bold{s}_i|^2}{w^2}}\bigg\{1-\frac{4}{w^2}\bold{s}_i^T\bold{d}+\frac{1}{2}\bold{d}^T\Big(\frac{16}{w^4}\bold{s}_i\bold{s}_i^T-\frac{4}{w^2}\bold{I}\Big)\bold{d}\bigg\}.
\end{aligned}
\end{equation}
By substituting~(\ref{dbeamapprox}) into~(\ref{jointmomentequation}), we get
\begin{equation}
\label{jointmomentproof}
\begin{aligned}
&E[Z_{m_1}Z_{m_2}{\cdots}Z_{m_{n_0}}]\\
&=\int_0^{2\pi}\int_0^{\infty}\prod_{i=1}^{n_0}\bigg[A_0e^{-\frac{2s_{m_i}^2}{w^2}}\bigg(1-\frac{4s_{m_i}}{w^2}\delta\cos(\phi_{m_i}-\theta)\\
&{\quad}+\frac{8s_{m_i}^2}{w^4}\delta^2\cos^2(\phi_{m_i}-\theta)-\frac{2}{w^2}\delta^2\bigg)\bigg]\cdot\frac{\delta}{2\pi{\sigma_s^2}}e^{-\frac{\delta^2}{2\sigma_s^2}}\,d{\delta}\,d{\theta}.
\end{aligned}
\end{equation}
With respect to the Rayleigh distributed $\delta$,~(\ref{jointmomentproof}) can be interpreted as an expected value of the polynomial. Consequently, we transform this into the integral of the product of cosine functions with coefficients involving Rayleigh moments. After calculating the integral of cosine functions with respect to $\theta$ by~(\ref{cosintegral}), the moment of a joint-pointing loss can be expressed without integral operations as~(\ref{jointmoment}).

\bibliographystyle{IEEEtran}
		\bibliography{VT-2022-01863}

\end{document}